 \documentclass[final,3p,times,twocolumn]{elsarticle}
\usepackage{graphicx}
\usepackage{hyperref}
\usepackage{amssymb}
\usepackage{isotope}
\usepackage{booktabs}
\usepackage{subcaption}
\usepackage{epigraph}
\usepackage{threeparttable}
\usepackage{amsthm}

\journal{Nucl. Instr. Meth. Phys. Res. A}

\begin{document}

\begin{frontmatter}

%% Title, authors and addresses

\title{Beyond the acceptance limit of DRAGON: the case of the $\isotope[6][]{Li}(\alpha,\gamma)\isotope[10][]{B}$ reaction}

%% use the tnoteref command within \title for footnotes;
%% use the tnotetext command for the associated footnote;
%% use the fnref command within \author or \address for footnotes;
%% use the fntext command for the associated footnote;
%% use the corref command within \author for corresponding author footnotes;
%% use the cortext command for the associated footnote;
%% use the ead command for the email address,
%% and the form \ead[url] for the home page:
%%
%% \title{Title\tnoteref{label1}}
%% \tnotetext[label1]{}
%% \author{Name\corref{cor1}\fnref{label2}}
%% \ead{psaltisa@mcmaster.ca}
%% \ead[url]{home page}
%% \fntext[label2]{}
%% \cortext[cor1]{}
%% \address{Address\fnref{label3}}
%% \fntext[label3]{}

%% use optional labels to link authors explicitly to addresses:
%% \author[label1,label2]{<author name>}
%% \address[label1]{<address>}
%% \address[label2]{<address>}

\author[McMaster]{A. Psaltis}
\ead{psaltisa@mcmaster.ca}
\author[McMaster]{A. A. Chen}
\author[TRIUMF]{D. S. Connolly}
\author[TRIUMF]{B. Davids}
\author[JINA,Bordeaux]{G. Gilardy}
\author[Ohio]{R. Giri}
\author[CSM]{U. Greife}
\author[TRIUMF,UNBC]{W. Huang}
\author[TRIUMF]{D. A. Hutcheon}
\author[CSM]{J. Karpesky}
\author[TRIUMF]{A. Lennarz}
\author[McMaster]{J. Liang}
\author[CSM]{M. Lovely}
\author[Ohio]{S. N. Paneru}
\author[TRIUMF]{C. Ruiz}
\author[UBC]{G. Tenkila}
\author[TRIUMF, York]{M. Williams}

%%Affiliations

\address[McMaster]{Department of Physics and Astronomy, McMaster University, Hamilton, Ontario L8S 4M1, Canada}
\address[TRIUMF]{TRIUMF, 4004 Wesbrook Mall, Vancouver, British Columbia V6T 2A3, Canada}
\address[JINA]{Department of Physics, Joint Institute for Nuclear Astrophysics, University of Notre Dame, Notre Dame, Indiana 46556, USA}
\address[Bordeaux]{Centre d'\'Etudes Nucl\'eaires de Bordeaux Gradignan, UMR 5797 CNRS/IN2P3 - Universit\'e de Bordeaux, 19 Chemin du Solarium, CS 10120, F-33175 Gradignan, France}
\address[CSM]{Department of Physics, Colorado School of Mines, Golden, Colorado 80401, USA}
\address[UNBC]{Physics Department, University of Northern British Columbia, Prince George, BC V2N 4Z9, Canada}
\address[UBC]{Department of Physics and Astronomy, University of British
Columbia, Vancouver, British Columbia V6T 1Z4, Canada}
\address[Ohio]{Department of Physics, and Astronomy, Ohio University, Athens, Ohio 45701, USA}
\address[York]{Department of Physics, University of York, Heslington, York YO10 5DD, United Kingdom}

\begin{abstract}
%% Text of abstract
Radiative capture reactions play a pivotal role for our understanding of the origin of the elements in the cosmos. Recoil separators provide an effective way to 
study these reactions, in inverse kinematics, and take advantage of the use of radioactive ion beams. However, a limiting factor in the study of radiative capture reactions in inverse kinematics is the momentum spread of the product nuclei, which can result in an angular spread larger than the geometric acceptance of the separator. 
The DRAGON facility at TRIUMF is a versatile recoil separator, designed to 
study radiative capture reactions relevant to astrophysics in the A$\sim$10-30 
region. In this work we present the first attempt to study with DRAGON a reaction, 
$\isotope[6][]{Li}(\alpha,\gamma)\isotope[10][]{B}$, for which the recoil angular spread 
exceeds DRAGON's acceptance. Our result is in good agreement
with the literature value, showing that DRAGON can measure resonance strengths
of astrophysically important reactions even when not all the recoils enter the separator. 

\end{abstract}

\begin{keyword}
recoil separators \sep inverse kinematics \sep radiative capture \sep resonance strength
%% keywords here, in the form: keyword \sep keyword

%% MSC codes here, in the form: \MSC code \sep code
%% or \MSC[2008] code \sep code (2000 is the default)

\end{keyword}

\end{frontmatter}

%%
%% Start line numbering here if you want
%%
%\linenumbers

%% main text
\section{Introduction}
\label{sec:intro}

%Table of reactions (p,g),(a,g)/ scenario/ max momentum cone
\begin{table*}[t!]
\centering
\caption{List of some astrophysically important reactions in the $A= 7 - 24$ mass region. 
The $Q$--value of each reaction is presented in MeV, along with the astrophysical
scenario that it affects and the respective energy region of the Gamow window in the center of mass system
(E\textsubscript{cm}). The $\theta_{r,max}$ value for each reaction is the maximum angle 
the recoils can have in the Gamow window listed in the second column.
The reactions marked with $\star$ have been measured using 
the DRAGON recoil separator, but not necessarily in the Gamow window, and the reactions marked with $\dagger$ have an
angular cone greater than DRAGON's maximum acceptance $\theta_{DRAGON} = \pm$ 21~mrad. 
See text for details.}
    \begin{tabular}{cccc} \hline
\hline 
    \textbf{Reaction} & \textbf{Q value (MeV)}  & \textbf{Astrophysical Scenario} & $\mathbf{\theta_{r,max}}$\\
     ~ & \textbf{E\textsubscript{cm} (MeV)}  & \textbf{(nucleosynthesis process)} & \textbf{(mrad)}\\ \hline 
     $\isotope[7][]{Li}(\alpha,\gamma)\isotope[11][]{B}^{\dagger}$ & 8.664 & Core--Collapse Supernovae & $\pm$51\\
    ~ & 0.7--3.6 & $\nu$--process & ~ \\ \hline
    $\isotope[7][]{Be}(\alpha,\gamma)\isotope[11][]{C}^{\dagger}$ & 7.544 & Core--Collapse Supernovae & $\pm$44 \\
    ~ & 0.5--1.2 & $\nu p$--process & ~ \\ \hline
    $\isotope[7][]{Be}(p,\gamma)\isotope[8][]{B}$  & 0.136 & Sun & $\pm$3 \\
    ~ & 0.02 & \emph{pp}--chains (solar $\nu$) & ~\\\hline
    $\isotope[12][]{C}(\alpha,\gamma)\isotope[16][]{O}^{\dagger \star}$ & 7.162 & Intermediate mass/Massive stars & $\pm$138 \\
    ~ & 0.03 & Quiescent helium burning & ~ \\ \hline
    $\isotope[13][]{N}(p,\gamma)\isotope[14][]{O}$ & 4.626 & X--ray bursts & $\pm$15 \\
    ~ & 0.3--2.2 & hot CNO cycle& ~ \\ \hline
     $\isotope[15][]{O}(\alpha,\gamma)\isotope[19][]{Ne}$ & 3.528 & X--ray bursts & $\pm$21 \\
    ~ & 1.5--4.6 & hot-CNO cycle  & ~ \\ \hline
    $\isotope[16][]{O}(\alpha,\gamma)\isotope[20][]{Ne}^{\dagger \star}$ & 4.730 & Intermediate mass/Massive stars & $\pm$87 \\
    ~ & 0.02 & Quiescent helium burning  & ~ \\ \hline
    $\isotope[17][]{O}(\alpha,\gamma)\isotope[18][]{F}^{\dagger}$ & 7.348 & AGB stars, massive stars, and novae & $\pm$64 \\
    ~ & 0.1--0.5& \emph{s}--process & ~ \\ \hline
    $\isotope[18][]{O}(\alpha,\gamma)\isotope[22][]{Ne}^{\dagger}$ & 9.667 & Intermediate mass/Massive stars& $\pm$55 \\
    ~ & 0.6--2.3  &  \emph{s}--process & ~ \\ \hline
    $\isotope[20][]{Ne}(\alpha,\gamma)\isotope[24][]{Mg}$ & 9.316 & Intermediate mass/Massive stars&  $\pm$105\\
    ~ & 0.04  &  Quiescent helium burning & ~ \\ \hline
    $\isotope[22][]{Ne}(p,\gamma)\isotope[23][]{Na}^{\star}$ & 8.794 &  AGB stars/ classical novae &  $\pm$18 \\
    ~ & 0.3--0.5  & Ne--Na cycle  & ~ \\ \hline
    $\isotope[22][]{Ne}(\alpha,\gamma)\isotope[26][]{Mg}^{\dagger}$ & 10.614 & Intermediate mass/Massive stars& $\pm$105 \\
    ~ & 0.038--1.450 &  \emph{s}--process & ~ \\ \hline
    $\isotope[23][]{Mg}(p,\gamma)\isotope[24][]{Al}^{\star}$ & 1.863 & O--Ne--Mg novae & $\pm$3 \\
    ~ & 0.5--0.9& Ne--Na cycle & ~ 
     \\  \hline \hline
    \label{tab:reactions}
    \end{tabular}
\end{table*}

Radiative capture reactions involving hydrogen and helium are of
pivotal importance for nuclear astrophysics. Knowing their cross 
sections improves reaction network calculations and thus our predictions
for the origin of the elements in the universe. Given that these reactions 
involve the two most abundant elements in the cosmos, they occur 
in almost any astrophysical scenario, including quiescent (\textit{e.g.} 
Ne--Na, Mg--Al cycles) and explosive (\textit{e.g.} $rp$--process, $\nu 
p$--process) stellar burning. 
These reactions are traditionally studied using intense proton and 
$\alpha$--beams from low--energy accelerators, impinging onto a heavy
target. This technique, even though it is still used until today 
with great success, has some drawbacks, such as the beam induced 
background and the inability to use short--lived targets.
Using recoil separators, radiative capture reactions can be studied
in inverse kinematics, with a heavy ion beam (stable or radioactive) 
impinging on a gas target (usually hydrogen or helium). Their advent remedies
the aforementioned problems, but imposed some new ones, mainly of a  geometric 
nature~\cite{ruiz2014recoil, brune2015radiative}.

The Detector of Recoils and Gammas of Nuclear reactions (DRAGON) 
facility in the Isotope Separator and Accelerator--I (ISAC--I) 
experimental hall at TRIUMF, Canada's particle accelerator centre 
in Vancouver, BC has carried out many of the radiative capture 
measurements involving radioactive ion beams to date. Even though it was 
constructed to study reactions with beams up to 
A=30~\cite{hutcheon2003dragon,vockenhuber2008improvements},  
over the last two decades, DRAGON has demonstrated versatility, 
having performed experiments from A= 3, \emph{e.g.},
$\isotope[3][]{He}(\alpha,\gamma)\isotope[7][]{Be}$~\cite{sjue2013beam},
to A= 76, \emph{e.g.}, $\isotope[76][]{Se}(\alpha,\gamma)\isotope[80][]{Kr}$~\cite{fallis2020first}.

The experiment presented in this work is a proof
of the capability of DRAGON to measure
resonant cross sections of radiative capture reactions 
of astrophysical interest in which the angular cone of 
the recoils exceeds its geometric acceptance. It was selected as a benchmark for the measurement of unknown resonance strengths of
the $\isotope[7][]{Be}(\alpha,\gamma)\isotope[11][]{C}$ 
reaction, which is important for $\nu p$--process nucleosynthesis 
(see Table~\ref{tab:reactions}). 
In the past, there have been acceptance--challenging experiments 
for nuclear reaction studies with DRAGON, such
as the $\isotope[12][]{C}(\isotope[16][]{O},\gamma)\isotope[28][]{Si}$ and $\isotope[12][]{C}(\alpha,\gamma)\isotope[16][]{O}$, reported in 
References~\cite{lebhertz201212, matei2006measurement}, but these studies did not 
involve the measurement of a resonance strength.
For this test we used a known resonance at a center of mass energy of
$E_{r}=$ 1458.5(6)~keV of the $\isotope[6][]{Li}(\alpha,\gamma)\isotope[10][]{B}$ 
reaction whose strength was originally measured by Forsyth \emph{et 
al.} in forward kinematics~\cite{forsyth19666li}. 
The present measurement was performed in inverse kinematics, 
using a stable $\isotope[6][]{Li}$ beam provided by the TRIUMF
Off-Line Ion Source (OLIS)~\cite{jayamanna2008off}. 

The paper is structured as follows: in Section~\ref{sec:radiative} 
we give a brief overview of conducting experiments using recoil 
separators and the challenges of studying reactions with low mass 
ion beams. In Section~\ref{sec:previous} we discuss the 
measurement by Forsyth \emph{et al.}. In 
Section~\ref{sec:experiment} we give an overview of DRAGON and the 
experimental setup, in Section~\ref{sec:analysis} we present the 
data analysis and the results, and finally in 
Section~\ref{sec:discussion} we present our conclusions by 
discussing the final results in more detail.
\section{Radiative capture reactions using recoil separators}
\label{sec:radiative}
Radiative capture reactions in inverse kinematics occur 
in a usually gaseous target, at rest in the laboratory frame, 
with the entire laboratory momentum being carried by 
the beam. The compound nucleus is formed in an excited state with energy
\begin{equation}
\label{eq:1}
    E_x = E_{cm} + Q
\end{equation}
where $E_{cm}$ is the energy in the center of mass system, and when the reaction
proceeds through a resonance, $E_{cm}=E_r$. $Q=(m_1+m_2-m_3)c^2$ is the 
reaction $Q$ value, where $m_1,m_2$ and $m_3$ are the masses of the projectile, 
the target, and the recoil, respectively.
%
%Figure of radiative capture
\begin{figure}[ht!]
\centering\includegraphics[width=\linewidth]{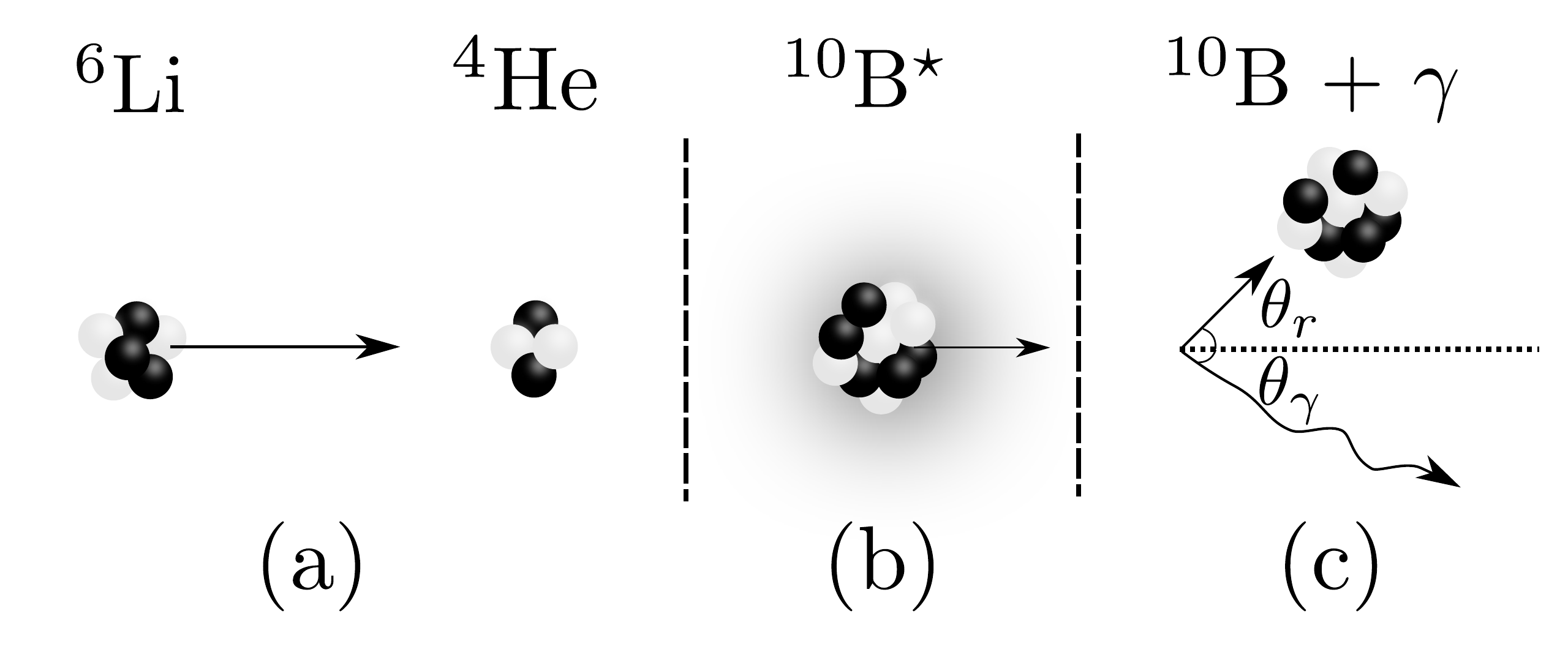}
\caption{Schematic representation of a radiative $\alpha$--capture on $\isotope[6][]{Li}$
in inverse kinematics: (a) beam ($\isotope[6][]{Li}$) and target ($\isotope[4][]{He}$) 
particles interact, (b) the compound nucleus ($\isotope[10][]{B}$) is synthesised in an 
excited state and then, (c) it decays by emitting a $\gamma$ ray. The recoil nucleus and the 
$\gamma$ ray are emitted in angles $\theta_r$ and $\theta_\gamma$ in the lab system respectively. 
See the text for details.}
\label{fig:cone}
\end{figure}

The excited nucleus decays by emitting one or multiple $\gamma$ rays
($\sum_i E_{\gamma_i}= E_x$),
which carry some of the initial momentum, and thus the products (recoil nuclei) 
form a narrow cone centered on the beam direction (see Figure~\ref{fig:cone}). 
In the simple case of a single $\gamma$ transition to the ground state, 
perpendicular to the beam direction ($\theta_\gamma = \pi/2$), the maximum angle of
the recoil nucleus can be calculated to be:
\begin{equation}
\label{eq:2}
\theta_{r,max} \simeq \arctan \left( \frac{Q+E_{cm}}{\sqrt{2 m_1c^2 \left( \frac{m_1+m_2}{m_2}\right) E_{cm}}} \right)
\end{equation}
A similar relation to  Equation~\ref{eq:2} can also be derived for the momentum spread 
of the recoils, $\Delta p /p$. 
The minimum of both relations appears at $E_{cm}=Q$. 
This behaviour is very interesting, since in the 
astrophysically relevant energy region, reactions with $Q/E_{cm} < 1 $ have 
increasing $\theta_{r,max}$ with 
increasing energy, while reactions with $Q/E_{cm} > 1$
exhibit the opposite behaviour. For example, the $\isotope[7][]{Be}(p,\gamma)\isotope[8][]{B}$
reaction, with a \emph{Q} value of 136.4~keV can be a very challenging
measurement for resonances with $E_r > Q$, since $\theta_{r,max}$ increases
with increasing energy~\cite{ruiz2014recoil}. On the other hand, for resonances with
$E_r \sim 1$~MeV, typical for astrophysical environments, the 
$\isotope[6][]{Li}(\alpha,\gamma)\isotope[10][]{B}$ (\emph{Q}-value = 4461.19~keV), has a decreasing 
$\theta_{r,max}$ with increasing energy, which is similar to the $\isotope[7][]{Be}(\alpha,\gamma)\isotope[11][]{C}$
reaction (\emph{Q}-value = 7543.6~keV), and for this reason is a good choice for a surrogate reaction.
For a more detailed discussion regarding the kinematics 
formalism of radiative 
capture reactions using recoil separators for astrophysics 
the reader is referred to
References~\cite{ruiz2014recoil, brune2015radiative}.

\begin{figure*}[!ht]
\centering
        \includegraphics[width=.49\textwidth]{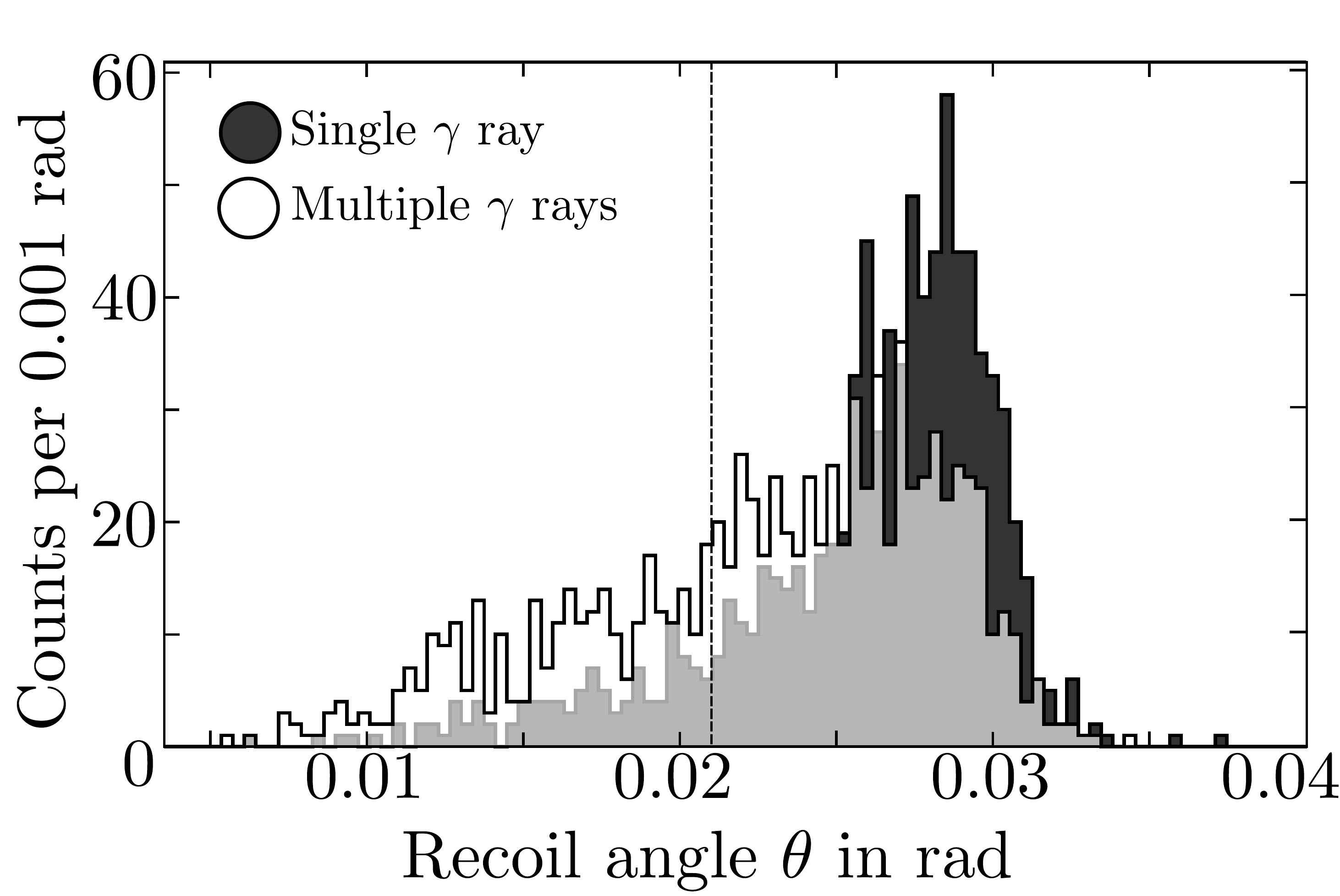}
        \includegraphics[width=.49\textwidth]{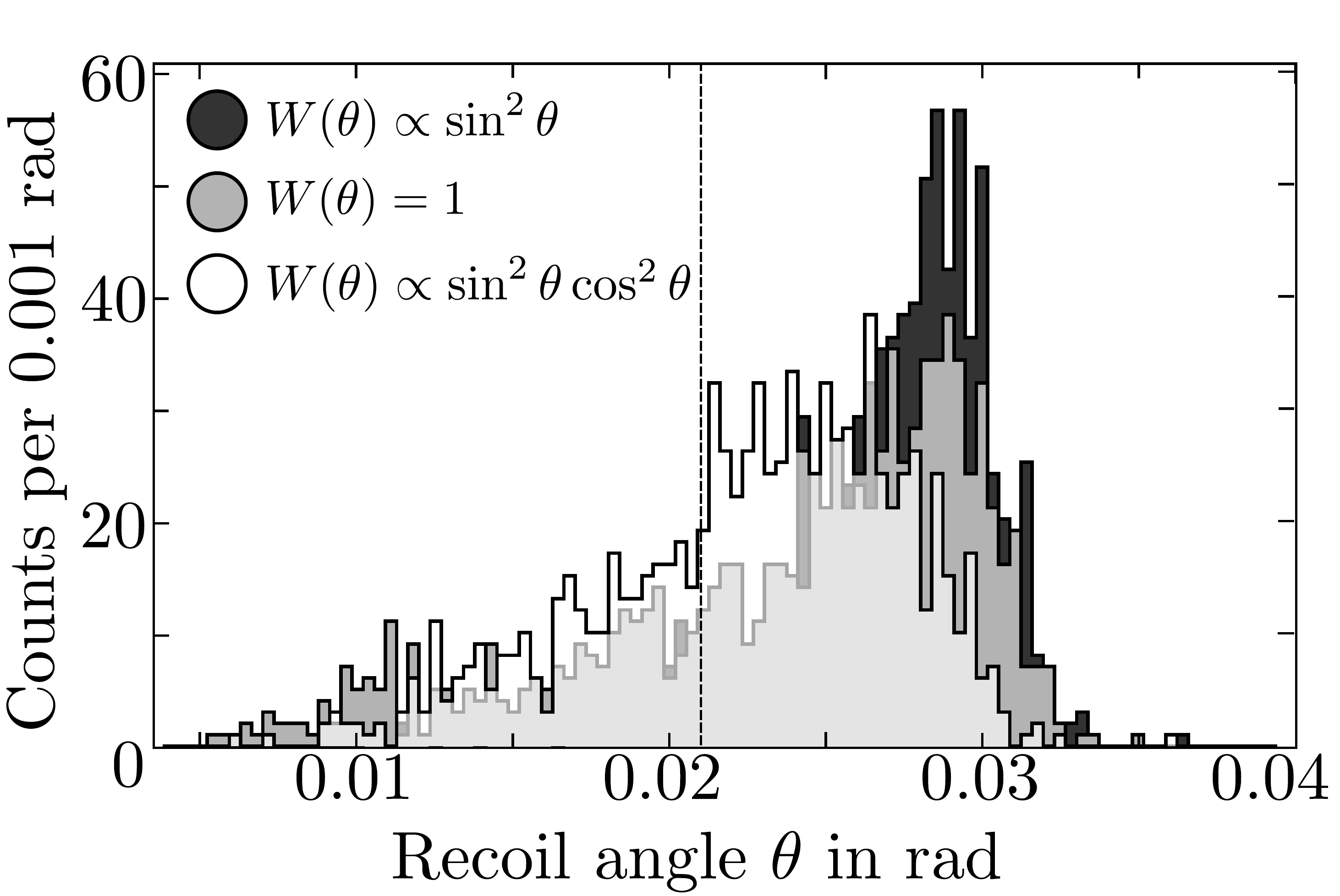}
        \caption{\texttt{GEANT} simulation results for the recoil angular distribution
        of the $E_{r}=$ 1458.5 keV resonance of $\isotope[6][]{Li}(\alpha,\gamma)\isotope[10][]{B}$ reaction by changing
        (\textit{Left}) the number of $\gamma$ rays emitted in the cascade and (\textit{Right}) their angular distribution (\textbf{Dipole} - $W(\theta) \propto \sin^2\theta$, \textbf{Uniform} - $W(\theta) = 1$ and \textbf{Quadrupole} - $W(\theta) \propto \sin^2\theta \cos^2\theta$.)
        The vertical line shows DRAGON's angular acceptance, $\pm$21~mrad. See the text for details.}
        \label{fig:geant}
\end{figure*}

What is really important in the case of an experimental study is not the 
maximum cone angle of the recoils, but rather their angular distribution, 
which affects the number of recoils within a given angular range of zero 
degrees, ultimately defining the transmission efficiency of recoils through the separator.
The recoil angular distribution 
depends on the $\gamma$ cascade and more specifically on the $\gamma$ 
branching ratios and the $\gamma$ angular distribution. 
To illustrate the above statements, we show in Figure~\ref{fig:geant} how the
recoil angular distribution of the $E_{r}=$ 1458.5(6) keV resonance of 
$\isotope[6][]{Li}(\alpha,\gamma)\isotope[10][]{B}$ reaction can be affected 
by changing either the number of the $\gamma$ rays in the cascade (left) or their 
angular distribution (right), which can be a M1/E2 decay
from the $J= 2^+$ state to the $J= 1^+$ and $J=3^+$ states. 
It is evident that a single transition to the 
ground state results in a distribution with a peak closer to the maximum angle $\theta_{r,max}$, 
while multiple $\gamma$ rays shift the distribution to smaller angles. 
This behaviour affects both 
the recoil transmission through the separator and the efficiency of the $\gamma$ ray detection system. 
As far as the $\gamma$ angular correlations are concerned, for the case
of radiative $\alpha$ capture on $\isotope[6][]{Li}$ in the center of mass system, 
the quadrupole distribution ($W(\theta) \propto \sin^2\theta 
\cos^2\theta$) shifts the average recoil momentum angle to smaller
angles, compared to the uniform ($W(\theta) = 1$) and dipole 
($W(\theta) \propto \sin^2\theta$) cases, as we can see in 
Figure~\ref{fig:geant} -- Right. Table~\ref{tab:fig1} 
shows an overview of the \texttt{GEANT} simulation that we discussed in the above.

\begin{table*}
\centering
    \caption{Tabulated results of the \texttt{GEANT} simulations presented in Figure~\ref{fig:geant}. See the text for details.}
    \begin{tabular}{ccccc} \hline \hline
      \textbf{\# of emitted} & $\mathbf{\gamma}$ \textbf{angular} & \textbf{Separator} & \textbf{BGO efficiency} \\ 
       $\mathbf{\gamma}$~\textbf{rays} & \textbf{distribution,} $\mathbf{W(\theta)}$  & \textbf{transmission (\%)} & \textbf{(\%)} \\ \hline \hline 
        1 & Uniform & $5.89 \pm 0.83$ &  $73.58 \pm 15.52$\\
        3 & Uniform & $27.4 \pm 1.9$  &  $80.4 \pm 7.6$ \\ \hline 
        2 & Uniform & $7.31 \pm 0.93$ &  $69.7 \pm 13.4$\\
        2 & Dipole & $11.86 \pm 1.21$ &  $64.81 \pm 9.95$\\
        2 & Quadrupole & $17.08 \pm 1.48$ &  $73.25 \pm 8.99$\\ \hline \hline
    \end{tabular}
   
    \label{tab:fig1}
\end{table*}

Recoil separators are built with an intrinsic angular acceptance, which sets 
a geometric limit to the number of reactions they can study. 
Table~\ref{tab:reactions} shows 
an overview of some important astrophysical reactions and their respective recoil cone angles
at energies relevant for astrophysics. For some of them the
maximum momentum angle of the recoils is quite large ($>$30~mrad), posing 
a great challenge to study them in inverse kinematics using recoil separators. The reaction we
selected to study in this work, $\isotope[6][]{Li}(\alpha,\gamma)\isotope[10][]{B}$, has a maximum 
recoil angle of $\theta_{r,max}= \pm 32$~mrad at $E_{cm}=$ 1458.5(6)~keV, which is 22~mrad greater that {DRAGON}'s angular 
acceptance ($\theta_{DRAGON}= \pm 21$~mrad).

In cases like that, the planning of an experiment and the subsequent analysis
relies heavily on detailed simulations of the separator (\texttt{GEANT} in the case of DRAGON) which provides 
information about the transmission of the recoils and the resonance energy. It is
very useful to have a prior knowledge of the $\gamma$ branching ratios and the
$\gamma$ angular distributions, but even in the case of a completely unknown 
$\gamma$ cascade, simulations can be used to estimate the branching ratios~\cite{ruiz2014recoil}.

\section{Previous Measurement}
\label{sec:previous}

The only published measurement of $\isotope[6][]{Li}(\alpha,\gamma)\isotope[10][]{B}$ 
reaction's $E_{r}=$1458.5(6) keV ($E_x$= 5919.5(6) keV) resonance strength was 
performed by Forsyth \emph{et al.}~\cite{forsyth19666li}. The
measurement was carried out at the University of Maryland Van de 
Graaff accelerator lab in regular kinematics, using a singly--charged
$\isotope[4][]{He}$ beam ($E_\alpha$= 0.9--3.3 MeV \& $I_\alpha= 2.5~\mu A$) 
and a 96\% isotopically enriched $\isotope[6][]{Li}$ target. $\gamma$ rays were detected 
using a NaI crystal placed at $90^{\circ}$ with respect to the beam. 

\begin{figure}[ht!]
\centering\includegraphics[width=0.9\linewidth,  clip=true, trim=140 0 140 0]{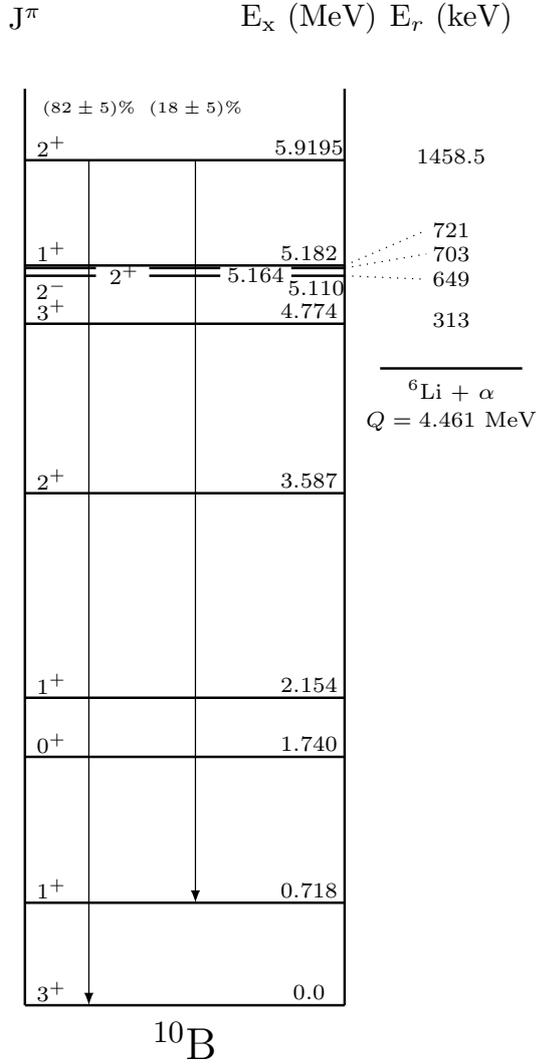}
\caption{Partial level scheme of the low--lying energy levels of 
$\isotope[10][]{B}$. Two $\gamma$--transitions of the $E_x$= 5919.5(6) keV state are shown. The reaction $Q$ value was taken from AME2016~\cite{wang2017ame2016}.}
\label{fig:scheme}
\end{figure}

The resonance strength was found to be $\omega\gamma= 0.228(38)$ eV 
and its width $\Gamma= 6(1)$~keV in the center of mass system.
Branching ratios of the $\gamma$ transitions were determined to be $82(5)~\%$
and $18(5)~\%$ to the ground state and the first excited state, respectively 
(see Figure~\ref{fig:scheme}), contrary to a single transition to the ground 
state reported in a study by Meyer--Sch\"utzmeister and
Hanna~\cite{meyer1957energy}. 
The reported branching ratios were used as input for the \texttt{GEANT} 
simulations of DRAGON, which provided the recoil transmission and the BGO $\gamma$ array detection efficiency (see Section~\ref{sec:lla}).

%Figure with level scheme of 10B

\section{Experimental Details}
\label{sec:experiment}

Our study was carried out in inverse kinematics using a beam of 
$\isotope[6][]{Li}^{+}$ from {OLIS}, which was accelerated 
through the ISAC--I Radio--Frequency Quadrupole (RFQ) and Drift--Tube Linac 
(DTL) to an average energy of 0.612(1)~A MeV ($E_{lab}$= 3.675(6)~MeV, 
$E_{cm}$= 1.468(3)~MeV), so that the resonance was centered in the gas 
target. The beam energy spread was $\Delta E / E \leq 0.3~\%$ throughout the 
experiment~\cite{laxdal2003acceleration}, with an average intensity 
of $1.94 \times  10^{10}~ s^{-1}$ (see also Section~\ref{subsec:beamnorm}). 
The windowless gas target pressure was maintained at P= 5.0(1) Torr,
corresponding to a thickness of $1.97(4) \times 10^{18}$ atoms/$cm^2$.
Choosing the aforementioned beam energy and gas target pressure, we were covering
a center--of--mass energy window of $1458.5 \pm 10$~keV.
The most intense charge state of the recoils ($\isotope[10][]{B}^{2+}$) 
was tuned through the separator to a 66~$\mu m$ thick, gridded 
Double--Sided Silicon Strip Detector ({DSSSD}) placed near the focal 
plane of {DRAGON} with a typical rate of 15--20~Hz.

\section{Data Analysis \& Results}
\label{sec:analysis}

To extract the resonance strength and compare it to the literature value, 
we first had to calculate the reaction yield, which includes 
identifying the recoils, determining the total number of beam 
particles, measuring the charge--state fraction of the recoils, and 
calculating the efficiency of the {BGO} array as well as the 
transmission of the recoils through the separator 
using \texttt{GEANT} simulations.

\subsection{Particle Identification}
\label{subsec:pid}

The $\isotope[10][]{B}$ recoils were detected by the DSSSD in 
coincidence with $\gamma$ rays in the BGO array.
Further discrimination was provided using software cuts on the 
separator time--of--flight (see Figure~\ref{fig:results1}), which is 
defined as the time difference between a $\gamma$ hit in a BGO 
detector and a hit in the DSSSD at the focal plane of DRAGON
~\cite{christian2014design}. DRAGON is very efficient in
rejecting unreacted beam ions for ($\alpha,\gamma)$
reactions, with demonstrated suppression factors of $>10^{13}$, which can be
increased by few orders of magnitude, by using the aforementioned software 
cuts~\cite{ruiz2014recoil, sjue2013beam,hutcheon2008background}.

\subsection{Beam Normalization}
\label{subsec:beamnorm}

To ensure a precise measurement of the reaction yield, we monitored
the beam current throughout the experiment using Faraday 
cups located along DRAGON. In particular, the number of beam
ions $N_{beam}$, impinging on the windowless gas target is
calculated using the following method: a silicon surface barrier
(SSB) detector placed at a well--defined lab angle of $57^\circ$ 
inside the target was detecting the elastically scattered gas target 
particles during each run. For a time window $\Delta t\sim$~240~s,
before and after each run, we recorded these measurements. 
At the same time, beam current measurements were made at a Faraday cup 
located 2~m upstream of the target. The normalized number 
of beam ions, $\mathrm{N_{beam}}$, is then given by:
\begin{equation}\label{eq:beam}
\mathrm{N_{beam} =  \mathcal{R} N_\alpha \frac{E^2}{P}}
\end{equation}
where $E$ is the beam energy and $P$ is the gas target pressure.
$\mathcal{R}$ is the normalization coefficient, given by:
\begin{equation}
\mathrm{\mathcal{R} = \frac{I}{q|e|} \frac{\Delta t}{N_\alpha} \frac{P}{E^2}\eta_{trg}}
\end{equation}
where $I/|e|$ is the current reading at the aforementioned Faraday Cup in ions per second,
$\eta_{trg}$ is the beam transmission through an empty target,
$q$ is the charge state of the beam ($1^+$), and
$\mathrm{N_\alpha}$ is the number of scattered target ($\alpha$) 
nuclei into the surface barrier detector during $\Delta t$.

\subsection{Boron charge state distribution}
\label{subsec:csd}

\begin{figure}[ht!]
\centering\includegraphics[width=.9\linewidth]{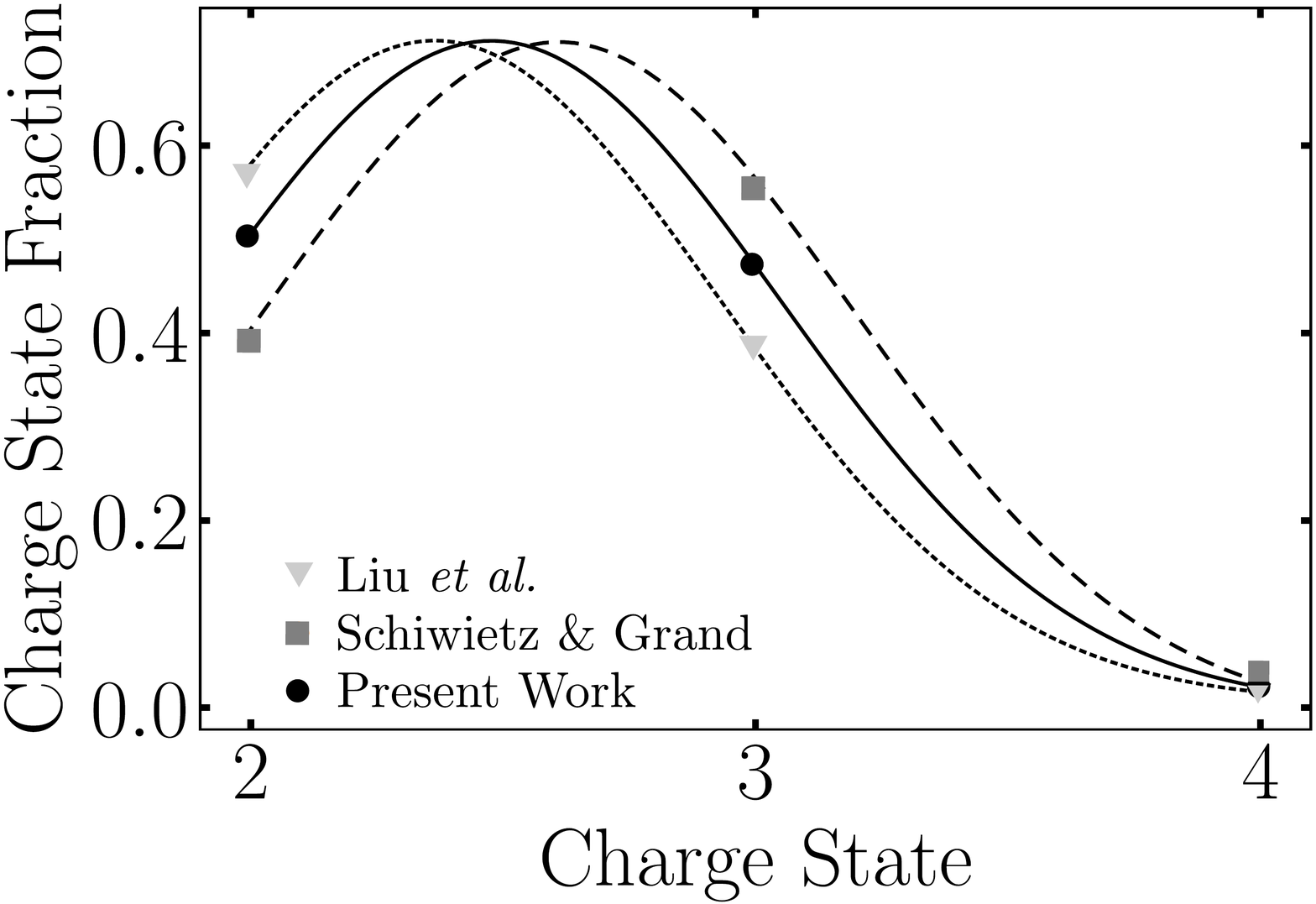}
\caption{Comparison between the experimentally measured CSD ($\eta_{CSD}$) for $\isotope[11][]{B}$ with the semi--empirical formulae of Liu \emph{et al.}~\cite{liu2003charge} and
Schiwietz \& Grand~\cite{schiwietz2001improved}. The fit to the experimental data is a Gaussian function. The error bars are smaller than the size of the points.}
\label{fig:csd}
\end{figure}

Given that {DRAGON} is tuned to select and transport only a single 
recoil charge state to the final focal plane, it is necessary to
measure the recoil charge state distribution (CSD) using a beam of an 
abundant isotope of the recoil element, to determine the total reaction 
yield. Charge State Distribution measurements were performed using a 
$\isotope[11][]{B}$ beam provided by {OLIS}. The results are compared 
with the semi--empirical formulae of Liu \emph{et al.}~\cite{liu2003charge} and
Schiwietz \& Grand~\cite{schiwietz2001improved} (see Figure~\ref{fig:csd}).
%Figure of CSD

\subsection{\isotope[6][]{Li} Stopping Power in \isotope[4][]{He}}
\label{subsec:stopping}

\begin{figure}[ht!]
\centering\includegraphics[width=\linewidth]{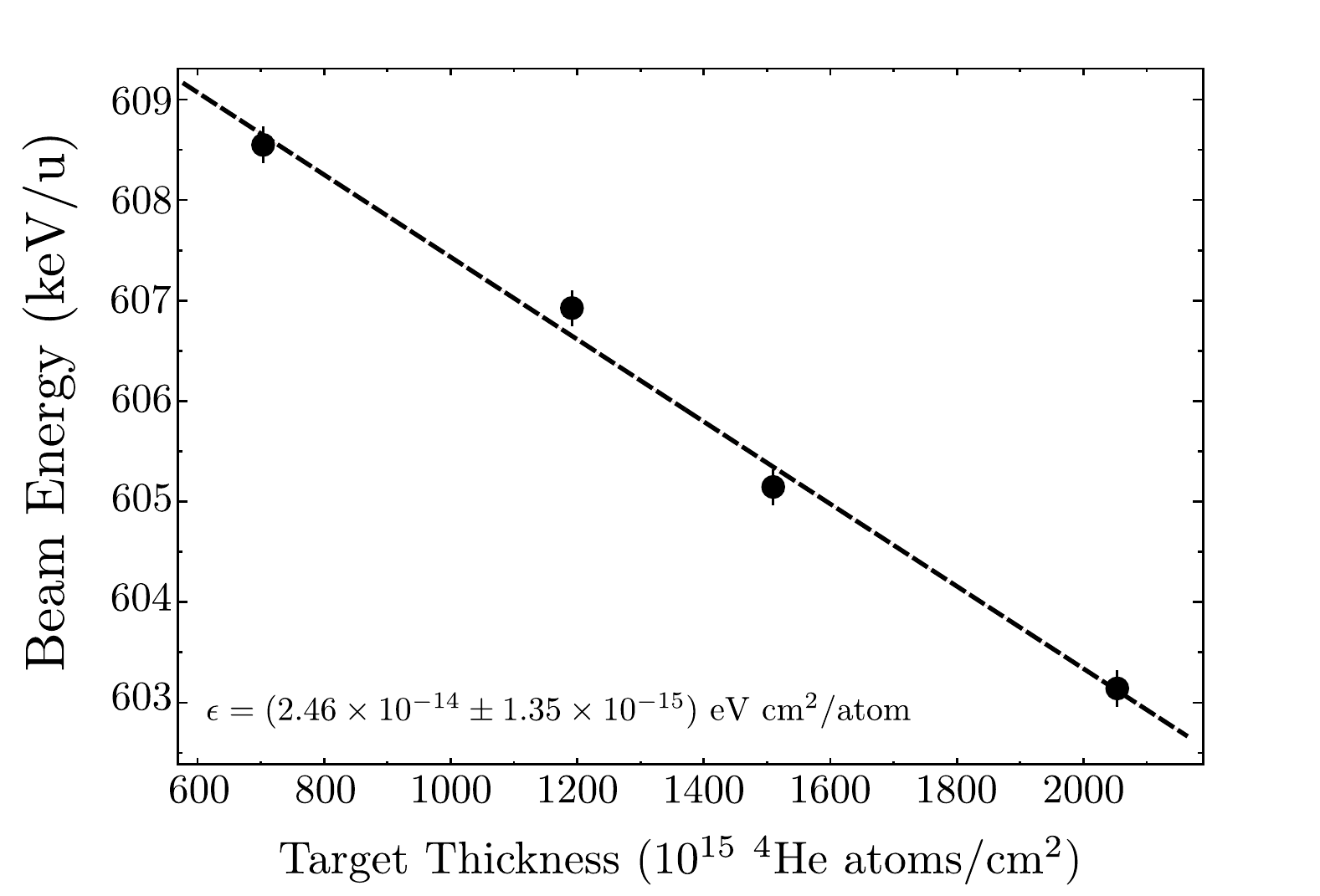}
\caption{Gas target thickness versus beam energy for different values of
    the target pressure. The slope of the linear fit is the stopping power $\epsilon$.}
\label{fig:stopping}
\end{figure}

One of the advantages of studying reactions using recoil separators 
is that the stopping power $\epsilon$, which is required for the calculation 
of the resonance strength, is measured directly and is
not based on semi--empirical formulae, which introduce an additional
uncertainty to the measurement, especially when they are 
extrapolated to low energies. At {DRAGON}, stopping powers are 
measured by varying both the pressure in the gas target
and the magnetic field strength needed to centre the beam
at a momentum dispersed angular focus in the 
focal plane of the first magnetic dipole of DRAGON (see Figure~\ref{fig:stopping}). We 
used these results to calculate the expected
yield $Y_{\omega\gamma_0}$ in Equation~\ref{eq:scaling} and
compare our experimental results with \texttt{GEANT} simulations 
(see Section~\ref{sec:lla}). 
%Figure of Stopping Power

\begin{figure*}[!ht]
\centering
        \includegraphics[width=.48\textwidth]{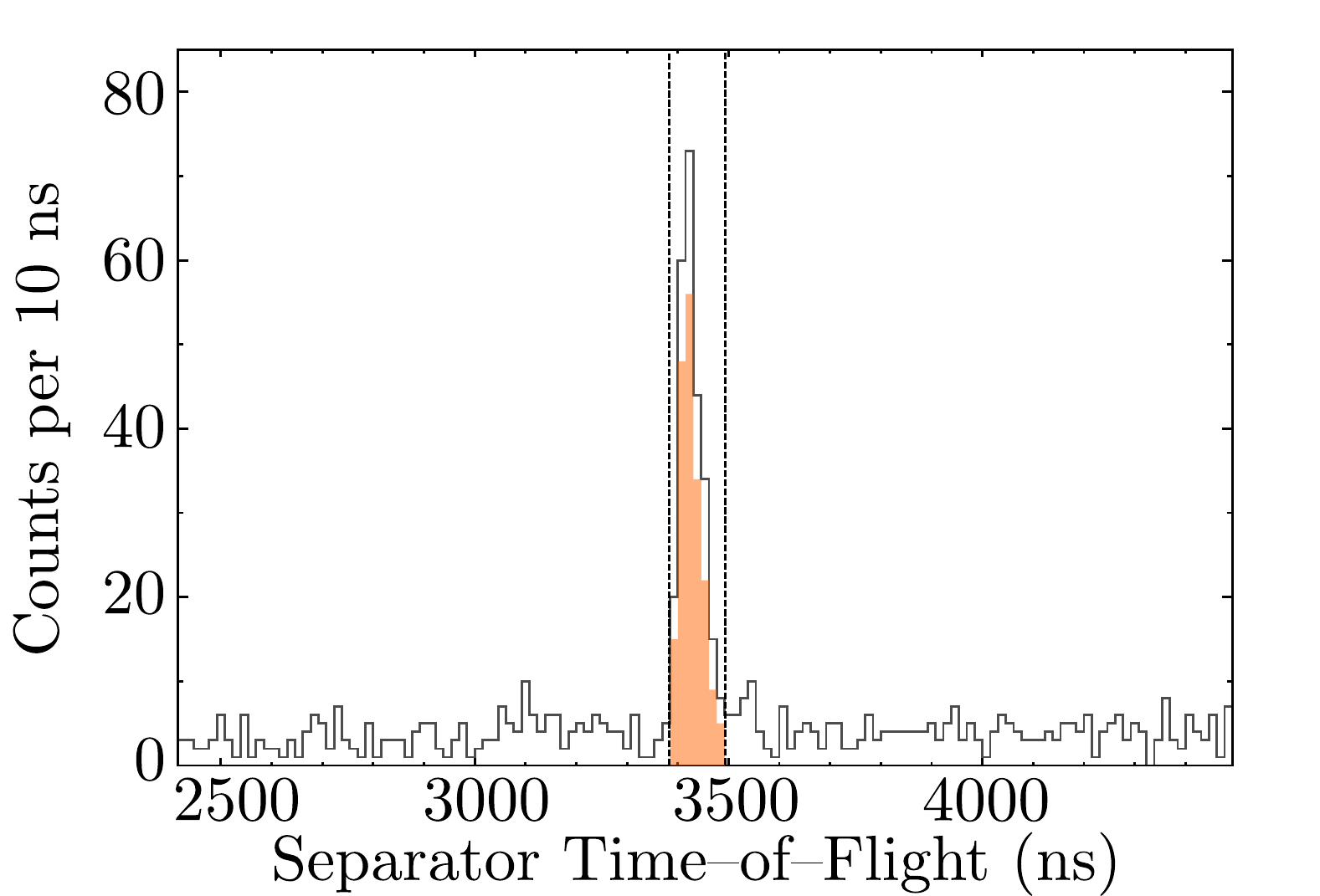}
        \includegraphics[width=.48\textwidth]{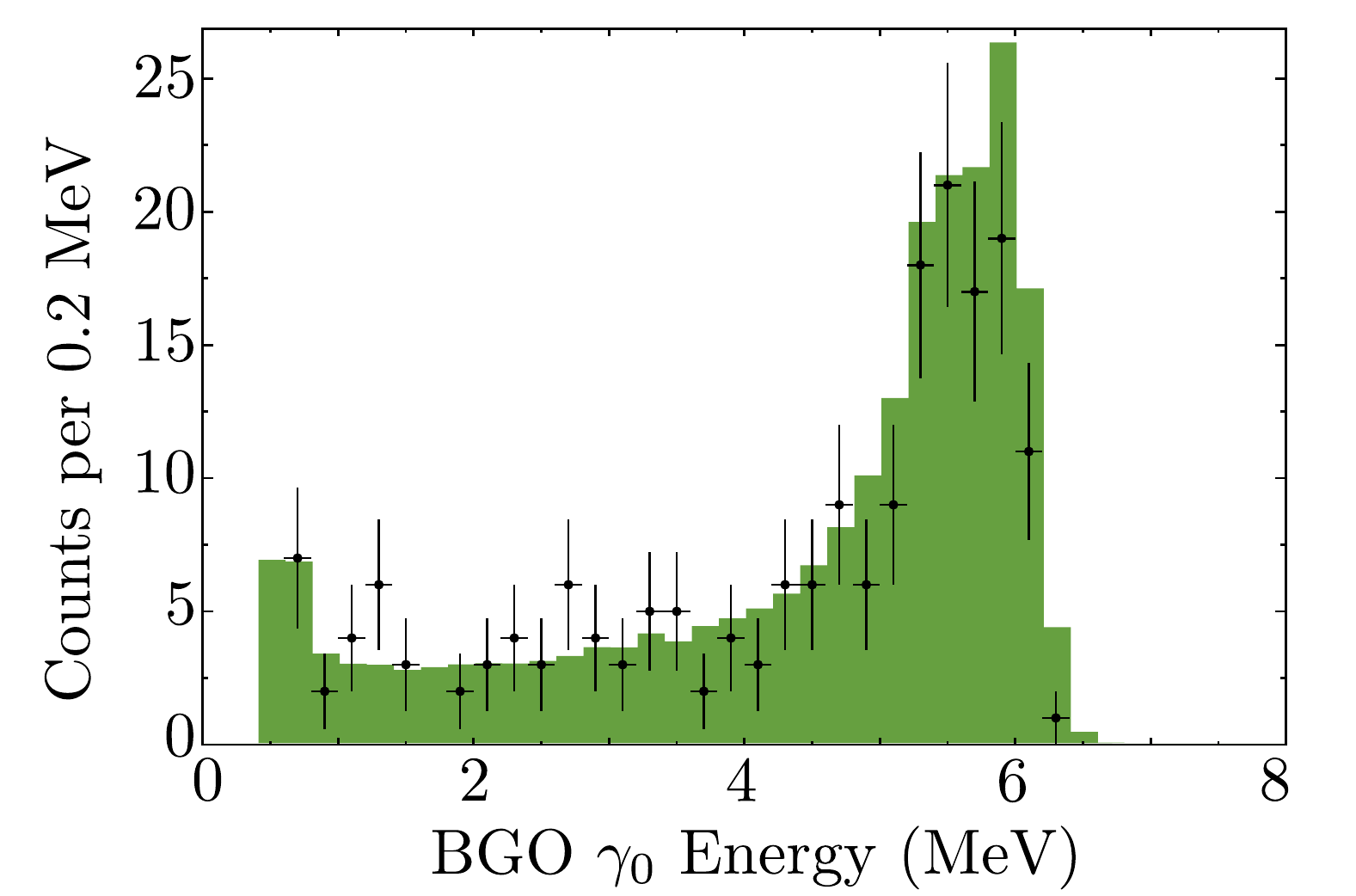}
        \includegraphics[width=.46\textwidth]{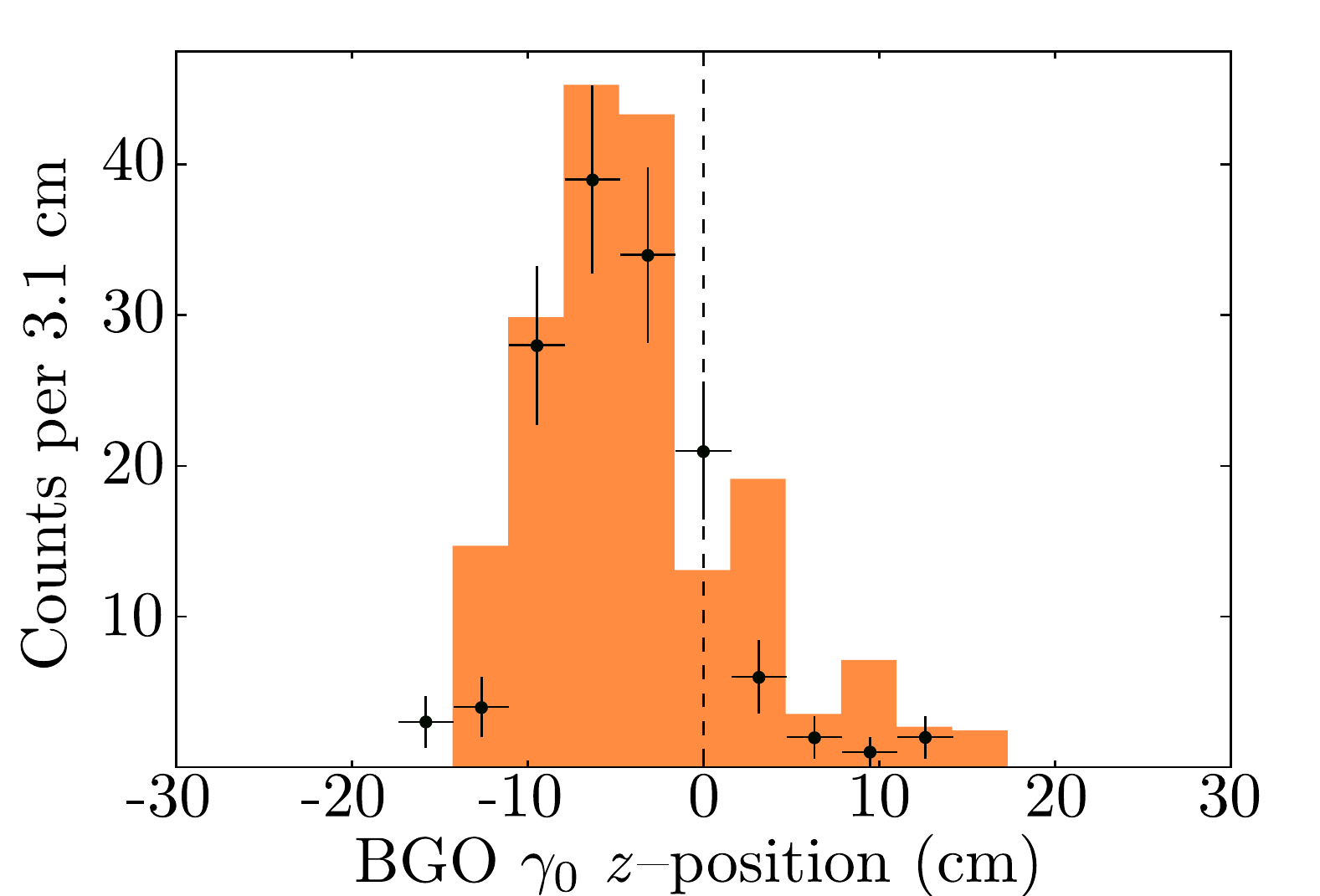}
        \hspace{0.01\textwidth}
        \includegraphics[width=.5\textwidth]{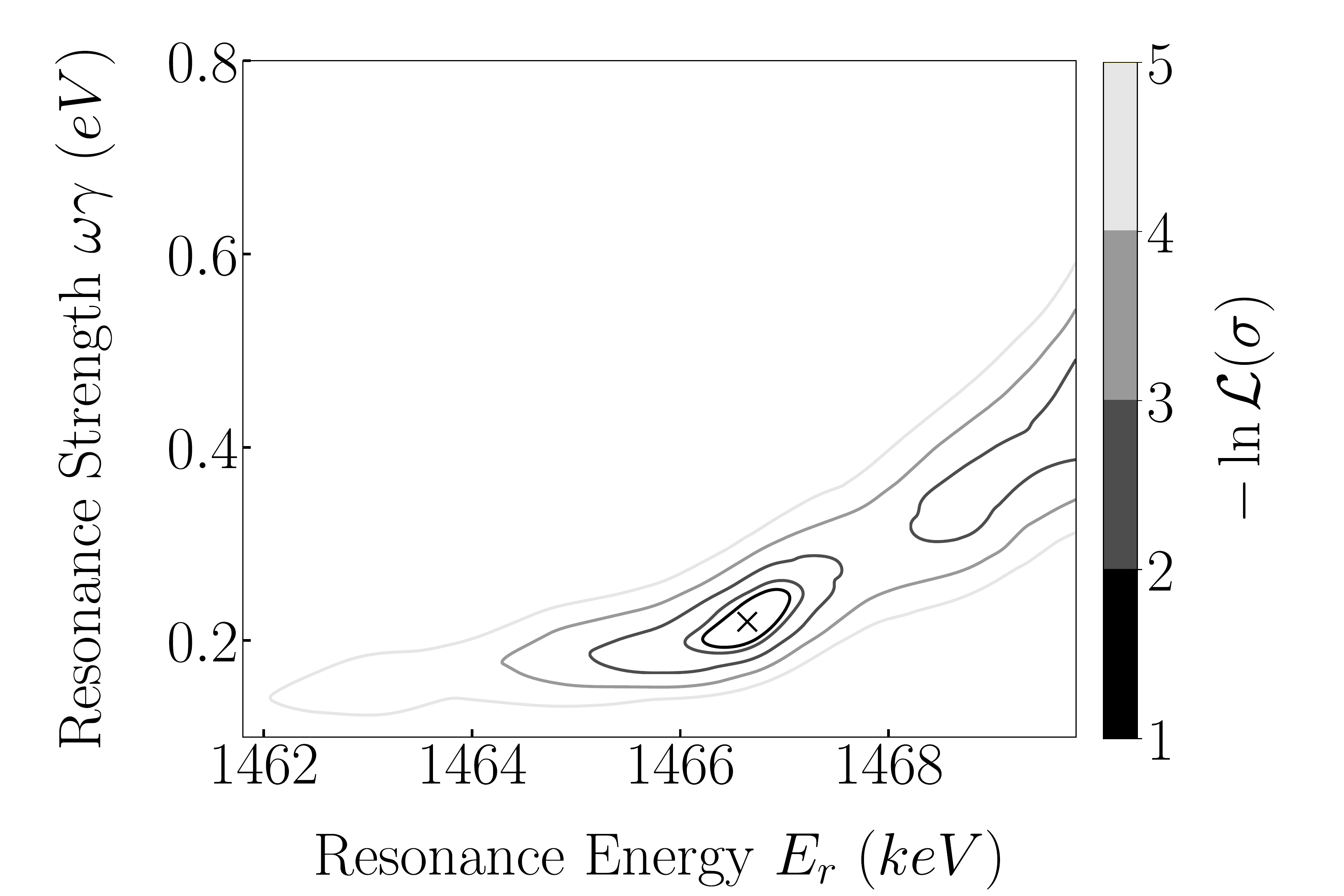}
        \caption{(Colour online) (Top Left) Separator time--of--flight spectrum for particle identification. The gate shown is for $\tau= 3.38-3.49~\mu s$. (Top Right) $\gamma_0$ energy plot comparison between experiment (black points) and \texttt{GEANT} simulation (green histogram). Both transitions of the $E_{r}=$1458.5 keV resonance can be seen.   (Bottom Left) The distribution of the z--position of the highest energy $\gamma$ ray for a yield measurement at P=5 Torr. The centroid is at -3.85~cm from the center of the gas target. The orange histogram shows the global best fit by means of \texttt{GEANT} simulations (Bottom Right) Negative Log--likelihood contour plot for the ($E_r, \omega \gamma$) space. The grey cross shows the minimum. See the text for a detailed discussion.}
        \label{fig:results1}
\end{figure*}

\subsection{Log-likelihood analysis for $E_r$ \& $\omega\gamma$}
\label{sec:lla}

The analysis of the BGO detector spectrum for the
highest energy $\gamma$ ray emitted by the de--excitation of the 
$\isotope[10][]{B}$ recoils (Figure~\ref{fig:results1})
shows that the resonance is excited upstream of the center of the gas 
target, indicating that 
the resonance energy is higher than 1458.5~keV. Therefore, we cannot use the 
standard method of {DRAGON} to determine the resonance energy from the 
distribution of the $z$ position of the highest energy $\gamma$ ray, 
since it assumes that the resonance is excited in the uniform density region 
surrounding the center of the gas target~\cite{hutcheon2012measurement}. 
Instead, we performed a likelihood analysis similar to the ones in
References~\cite{erikson2010first, christian2018direct} to extract the
resonance energy $E_r$, and its strength $\omega\gamma$.

To begin, we performed simulations for the $z$ distribution 
of the highest energy $\gamma$ rays using different resonance energies 
(13 values, spanning from $E_r$= 1457.8 to 1469.8 keV) and a fixed beam energy 
of 3.675~MeV and spread equal to the experimental one, 
with the standard {DRAGON} \texttt{GEANT3} simulation package\footnote{The
\texttt{GEANT3} simulation package of DRAGON can be found at 
\href{https://github.com/DRAGON-Collaboration/G3\_DRAGON}{\texttt{https://github.com/DRAGON-Collaboration/G3\_DRAGON}}}~\cite{gigliotti}. 
The \texttt{GEANT} input file included nuclear level information, such as
lifetimes and $\gamma$ branching ratios, from 
Reference~\cite{tilley2004energy}. For the $\gamma$ ray angular distribution,
which as we discussed earlier affects the recoil transmission through the 
separator, we proceeded as follows: a spin 1 beam ($\isotope[6][]{Li}$) on 
a spin 0 target ($\isotope[4][]{He}$) can populate $M=0, \pm 1$ magnetic
substates of $\isotope[10][]{B}$. Using Reference~\cite{rose1967angular},
we found the statistical tensor coefficients $\rho_2(2,0)=\rho_2(2,1)=-1.195$.
Then we multiplied with the geometry factors $R_2(2 \rightarrow 3)=0.1195$ and
$R_2(2 \rightarrow 1)=0.4183$, which results in the $\gamma$ angular 
distribution for the two transitions: 
$W(\theta)= 0.86 + 0.21 \sin^2(\theta)$ to the $J=3$ ground state and 
$W(\theta) = 0.5 + 0.75 \sin^2(\theta)$ to the $J=1$ first excited state.
Therefore the angular distribution of the 
dominant ground state transition is nearly isotropic and that of the other
is between isotropic and bipolar.

\begin{figure}[ht!]
\centering\includegraphics[width=\linewidth]{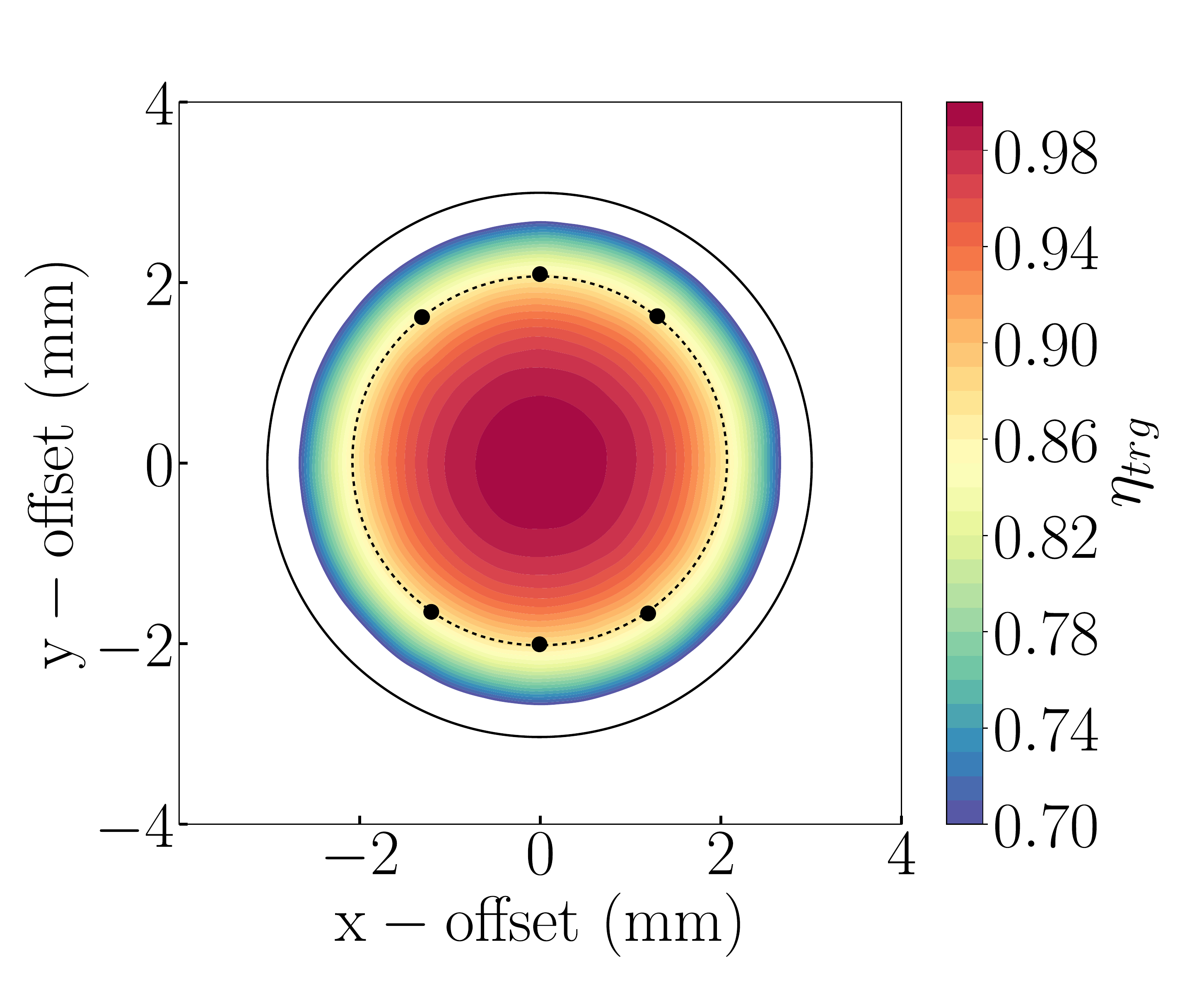}
\caption{(Colour online) Beam transmission results for a beam with 2$\times$rms size of 1.78~mm in x-- and y--axis. The black points show the locations selected for additional \texttt{GEANT} simulations. See the text for details.}
\label{fig:transmission}
\end{figure}

In addition, we took into account the transmission of the beam
through the gas target. In particular, during the experiment we measured an $\sim$86\% transmission of the
beam through the gas target by means of Faraday cup measurements. 
This implies that the beam was not centered as it was entering the gas target, given that
its $2 \times$rms size, assuming a Gaussian profile, was measured by the {ISAC} operators to be 1.78~mm in both x and y. To include this piece of 
information in our \texttt{GEANT} simulations, we performed Monte--Carlo simulations sampling both x-- and
y--axis offsets for the measured beam size. Figure~\ref{fig:transmission} shows the
results of the simulated transmission. After that, we selected six point with 86\% transmission to
perform our simulations (black points in Figure~\ref{fig:transmission}).

We then scaled the generated BGO spectra according to the expected 
reaction yield by a factor 
\begin{equation}\label{eq:scaling}
\eta\frac{Y_{\omega\gamma_0} N_{beam}}{N_{sim}},
\end{equation}
where $\eta$ is the recoil detection efficiency\footnote{It includes the recoil charge state fraction, the heavy ion detector efficiency, and the data acquisition dead time.}, $Y_{\omega\gamma_0}$ 
is the reaction 
yield from a single--level Breit--Wigner resonance ($\Gamma = 5.82$~keV~\cite{tilley2004energy}) of arbitrary 
strength $\omega \gamma$ (200 values spanning from 0.05 - 10 eV),  
$N_{beam}$ is the number of incident beam ions (See Section~\ref{subsec:beamnorm}) and $N_{sim}= 5 \times 10^4$ is the number of simulated events. The BGO array $\gamma$ ray detection 
efficiency $\eta_{BGO}$
and the separator transmission  $\eta_{separator}$ are built--in the \texttt{GEANT} simulation, and thus we do not include them in the $\eta$ factor. The simulated $\gamma$ spectra are convoluted with
a Gaussian resolution function with $\sigma(E) = 0.1733 \sqrt{E}/(\ln{2}\sqrt{8})$, which
is based on the experimentally measured resolution of the BGO array.
The scaled BGO spectra are then compared to the experimental data. From these simulations we created a ($E_r$, $\omega\gamma$)
space of 2,600 points ($13 \times 200$) and for each point on this grid, we calculated 
the negative log--likelihood using: 
\begin{equation}
\mathrm{-ln \mathcal{L}=  \sum_i  \left[ln(n_i!)-n_iln(f_i)\right] + S},
\end{equation}
where $i$ is the number of bins in the experimental BGO spectrum, 
$\mathrm{n_i}$ is the number of events in the $i^{th}$ bin,
$\mathrm{f_i}$ is the number of events in the scaled simulation
$i^{th}$ bin and $S$ the total number of events in the scaled histogram.

\begin{table}[ht!]
\centering
\caption{Settings of the \texttt{GEANT3} simulation for the Log--Likelihood
analysis. Nuclear properties were adopted from 
Reference~\cite{tilley2004energy}. See the text for details.}
\begin{tabular}{cc}
\hline \hline
\textbf{Quantity} & \textbf{Used Value} \\
\hline
 Excited state lifetime & $1.13 \times 10^{-19}$~s \\
 Resonance energy &   1457.8--1469.8~keV \\              
 Beam mass excess & 14.087~MeV\\
 Recoil mass excess & 12.051~MeV\\
 $\alpha$ partial width & $5.82$~keV\\
 $\gamma$ partial width & 0.1114~eV\\
 $\gamma$ branching ratios & 82\% (to the ground state) \\
                & 18\%  (to the first excited)\\
 $\gamma$ angular distributions  &    \\
 $2 \rightarrow 3$ (ground state) & $W(\theta) = 0.86 + 0.21 \sin^2(\theta)$  \\
 $2 \rightarrow 1$ (first excited) & $W(\theta) = 0.50 + 0.75 \sin^2(\theta)$ \\
\hline \hline
\end{tabular}
\label{tab:geant}
\end{table}

Figure~\ref{fig:results1} shows the results of our simulations with 
a single minimum for the negative log--likelihood, with energy
that corresponds to a location inside the gas target. The global minimum 
has $ -\ln \mathcal{L}_0 = 35.69$ and it is the only point where a 1$\sigma$ contour
can be deduced.

On top of the above analysis, we also performed tests on the 
\texttt{GEANT} simulation by changing the random seed of the Monte Carlo
simulation, to ensure that the distribution of events is Poissonian,
as in an experimental study. Due to the large number of simulation events 
($N_{sim}= 5 \times 10^4$) the final result does not depend on the random seed.

The sources of systematic uncertainty in the final result for the 
resonance strength are presented in Table~\ref{tab:uncert}. 
The most important source of systematic uncertainty is the BGO 
efficiency which accounts for 11.4\% and it was determined by
varying the $\gamma$ branching ratios of the resonance. 
The statistical uncertainty originates from the Log--Likelihood analysis
and is defined by the bounds of the 1$\sigma$ contour 
($\delta \omega \gamma = ^{+ 0.025}_{-0.035}$~eV)
For the level excitation energy, $E_x$, we have similarly taken into 
account the statistical uncertainty from the Log--Likelihood 
analysis, $\pm 0.5$~keV, and for the systematic 
uncertainty, we adopt the relative uncertainty of the beam energy 
$\delta E_{beam} = 0.16$\%, which yields $\delta E_x$(syst.)= 2.4~keV.

%Table relative uncertainties
\begin{table}[ht!]
\centering
\caption{Relative systematic uncertainties used to calculate the resonance strength of the $E_{r}=$1458.5 keV resonance.}
\begin{tabular}{lcc}
\hline \hline
\textbf{Quantity} & \textbf{Measured} & \textbf{Relative}\\
~ & \textbf{Value} & \textbf{Uncertainty} \\
\hline
 $\eta_{CSF}$ & $0.523(8)$ & 1.5\% \\
 $N_{beam}$ & $3.252(53) \times 10^{14}$ & 1.64\% \\
 $\eta_{BGO}$ & $0.332(38)$ & 11.4\% \\
 $\eta_{separator}$ & $0.122(5)$ & 4.1\% \\
 $\epsilon$ (eV cm$^2$) & $24.63(136) \times 10^{15}$ & 5.5\% \\
 $\eta_{livetime}$ & $0.91573(8) $ & 0.009\% \\
 $E_{beam}$ (A MeV) & $0.612(1)$   & 0.16\% \\
\hline
\textbf{Total systematic} &  ~  & 13.49\%\\
\textbf{uncertainty} & \\
\hline \hline
\end{tabular}
\label{tab:uncert}
\end{table}

The results for the resonance energy $E_r$, excitation energy $E_x$
and strength $\omega\gamma$, for this minimum are the following:
\[
    E_r =  1466.6~\pm 0.5~\textrm{(stat.)} \pm 2.4~\textrm{(syst.)}~\textrm{keV} 
\]
\[
    E_x =  5927.8~\pm 0.5~\textrm{(stat.)} \pm 2.4~\textrm{(syst.)}~\textrm{keV} 
\]
\[
    \omega \gamma = 0.225~^{+0.025}_{-0.035}~\textrm{(stat.)} \pm 0.030~\textrm{(syst.)}~\textrm{eV}
\]

\begin{figure*}[!ht]
\centering
        \includegraphics[width=.45\textwidth]{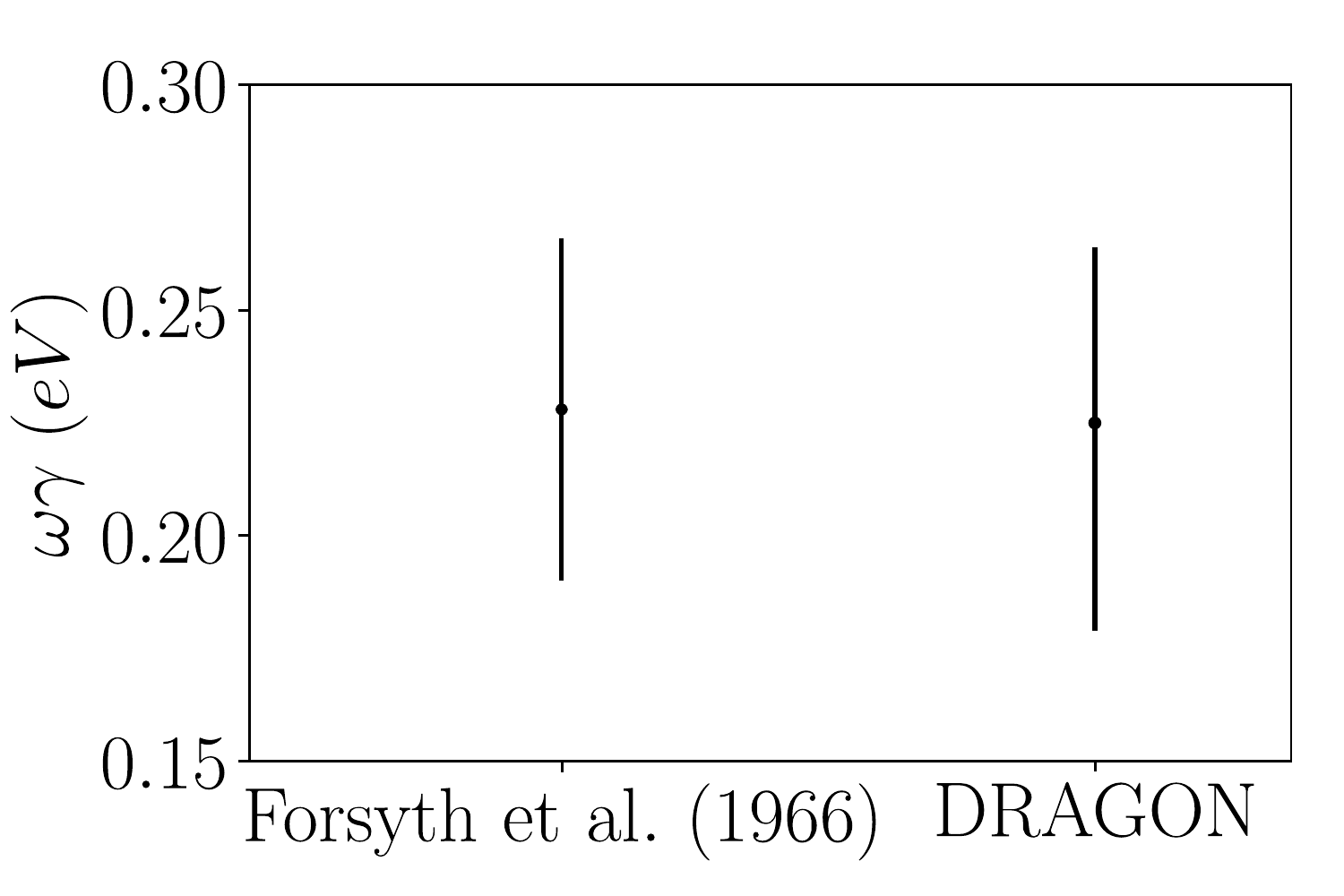}
        \includegraphics[width=.45\textwidth]{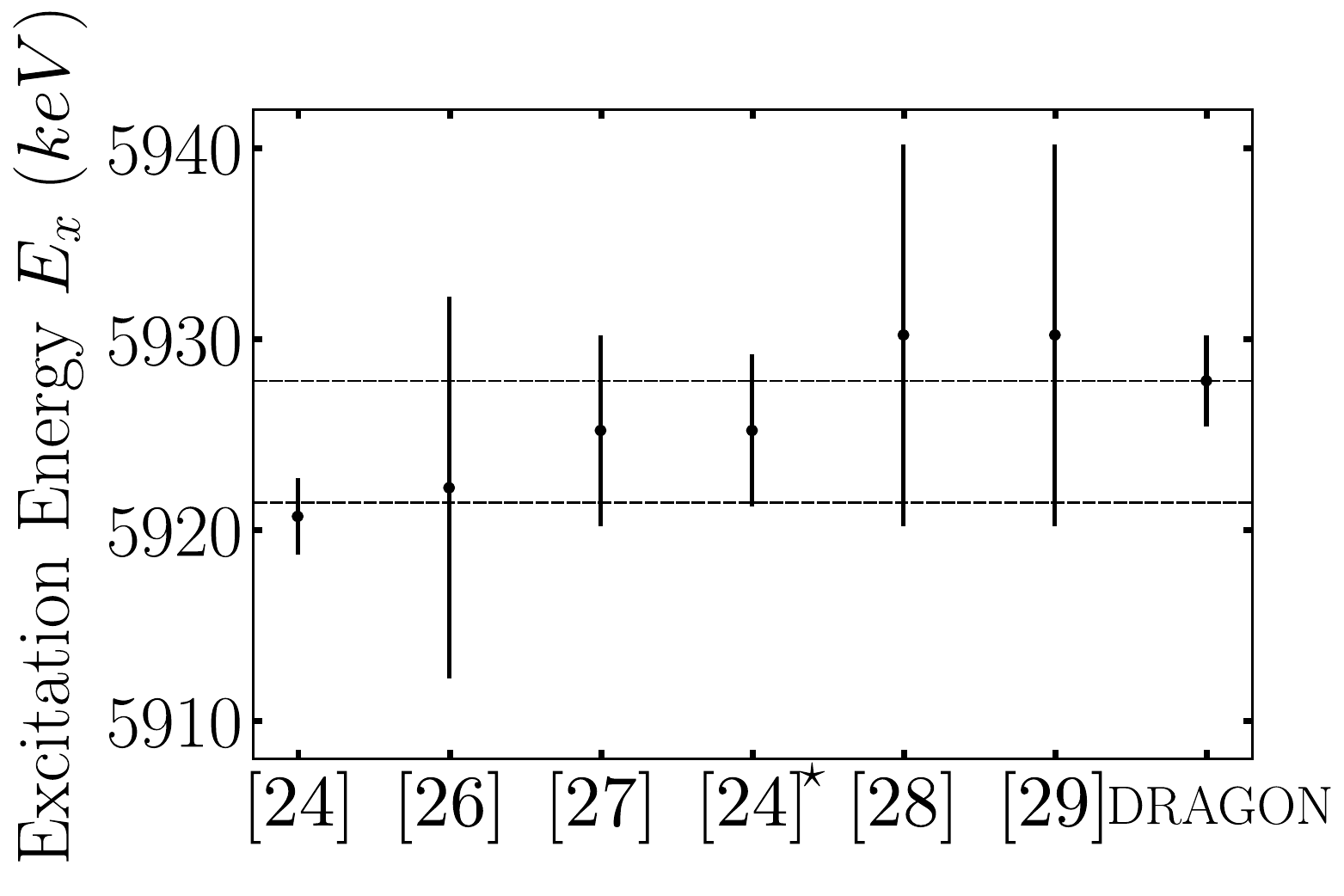}
        \caption{(Left) Comparison between the result of Forsyth \emph{et al.} and the present work for the resonance strength of the $E_r$= 1458.5~keV resonance. (Right) Excitation energies for the $E_x$= 5919.5~keV state of $\isotope[10][]{B}$ from normalized literature values (Table~\ref{tab:energies}) compared with the present work. The band correspond to the EVM average uncertainty. See the text for a detailed discussion.}
        \label{fig:results2}
\end{figure*}

Our final resonance strength is in excellent agreement with the measurement 
of Forsyth \emph{et al.}~\cite{forsyth19666li} as shown in 
Figure~\ref{fig:results2}. However, the resonance energy (excitation energy) we
extracted with {DRAGON} is higher than the recommended value 
in nuclear databases~\cite{nudat}. A literature search 
(References~\cite{kashy19749,buccino1965levels,armitage1964levels,gorodetzky1965study,park1973spectroscopy,park1973spectroscopy,fife1967excited}) shows 
that the excitation energy $E_x$, for the state of interest 
lies between 5920-5930~keV, which is consistent with
our result (see Table~\ref{tab:energies} and Figure~\ref{fig:results2}). 
To obtain an average literature--based excitation energy with a
realistic uncertainty, we first excluded 
the measurement of Buccino \& Smith~\cite{buccino1965levels} as 
an outlier ($E_x= 5900 \pm 80$~keV) applying Peirce's 
criterion~\cite{peirce1852criterion}.
We then calculated the average using the Expected Value Method (EVM)~\cite{birch2014method} for all the remaining measurements with 
reported uncertainties to be $E_x = (5920.3 \pm 2.4)$~keV. It is worth 
mentioning that the adopted
level energy of $E_x= (5919.5 \pm 0.6)$~keV~\cite{nudat}
is mainly determined by the high--precision
measurement reported in Reference~\cite{kashy19749}.
However, this result was extracted using an Enge split--pole 
spectrograph which usually results in excitation energy 
uncertainties of approximately $\pm 5$~keV~\cite{marshall2018focal}. 
For this reason, we also calculated the EVM average with a more realistic uncertainty 
for the study of Reference~\cite{kashy19749}, based on their analysis -- $\delta E_x= 2$~keV ---
to be $E_x= (5921.2 \pm 3.2)$~keV.
Furthermore, we investigated how the reaction $Q$ value changed between the literature measurements.
We found a $\sim 1-2$~keV difference between the $A=10$ evaluations and the AME2016 compilation, which we used for our calculations~\cite{wang2017ame2016, tilley2004energy,ajzenberg1959t, lauritsen1966energy, ajzenberg1974energy, ajzenberg1979energy, ajzenberg1984energy, AJZENBERGSELOVE19881} (see also
Table~\ref{tab:qval}). In particular, the \emph{Q} value changes between the 1979 to 1984 and 1984 to 1988
evaluations are based on mass measurements reported in References~\cite{chalupka1983precision} and~\cite{ellis1984precise}, respectively.
For this reason, we adjusted the literature values of the excitation energy to the current \emph{Q} value, assuming that they 
were following the latest $A=10$ evaluation at the time of publication (see Table~\ref{tab:energies}). 
The new result, $E_x= (5924.6 \pm 3.2)$~keV agrees within 1$\sigma$
with the energy extracted using the Negative Log--Likelihood 
analysis  $E_x= 5927.8~\pm 0.5~(stat.) \pm 2.4~(syst.)$~keV 
(see Figure~\ref{fig:results2}).

%Q-value evolution
\begin{table}[ht!]
\centering
\caption{Evolution of the $\isotope[6][]{Li}(\alpha,\gamma)\isotope[10][]{B}$ reaction $Q$ value through the years. See the text for details.}
\begin{tabular}{cccc}
\hline \hline
\textbf{Year} & $\mathbf{Q}$ \textbf{value (keV)} & $\mathbf{\Delta Q}$ \textbf{(keV)} & \textbf{Ref.}\\
\hline
 2017 & 4461.19 &  -    & ~\cite{wang2017ame2016}    \\
 2004 & 4461.10 & +0.09 & \cite{tilley2004energy}    \\
 1988 & 4459.60 & +1.59 & \cite{AJZENBERGSELOVE19881}\\
 1984 & 4460.30 & +0.89 & \cite{ajzenberg1984energy} \\
 1979 & 4460.50 & +0.69 & \cite{ajzenberg1979energy} \\
 1974 & 4460.00 & +1.19 & \cite{ajzenberg1974energy} \\
 1966 & 4461.00 & +0.19 & \cite{lauritsen1966energy} \\
 1959 & 4459.00 & +2.19 & \cite{ajzenberg1959t}      \\
\hline \hline
\end{tabular}
\label{tab:qval}
\end{table}

\begin{table*}
\centering
\caption{Summary of reported energies for the $E_x$= 5920 keV state of $\isotope[10][]{B}$ from different measurements, normalized to the current $\isotope[6][]{Li}(\alpha,\gamma)\isotope[10][]{B}$ 
reaction $Q$ value~\cite{wang2017ame2016}. See the text for a detailed discussion.}
    \begin{tabular}{cccc} \hline
\hline 
    \textbf{Reaction} & \textbf{E\textsubscript{x} (keV)}  & $\boldmath{\delta}$\textbf{E\textsubscript{x} (keV)}  & \textbf{Reference}\\ \midrule
     $\isotope[9][]{Be}(d,n)\isotope[10][]{B}$ &  5902.2 & 80  & Buccino \& Smith~\cite{buccino1965levels}\\
     $\isotope[10][]{B}(p,p^\prime)\isotope[10][]{B}$  &  5920.7  & 0.6 &  Kashy,  Benenson \& Nolen Jr.~\cite{kashy19749}\\
      $\isotope[10][]{B}(d,d')\isotope[10][]{B}$  &  5922.2  & 10 & Armitage \& Meads~\cite{armitage1964levels} \\
      $\isotope[10][]{B}(p,p')\isotope[10][]{B}$  &  5922.2  & 10 & Armitage \& Meads~\cite{armitage1964levels} \\
      $\isotope[11][]{B}(\isotope[3][]{He},\alpha)\isotope[10][]{B}$  &  5925.2  & 5 &  Gorodetzsky \emph{et al.}~\cite{gorodetzky1965study} \\
      N/A  &  5925.2  & 4 &  Reported in Reference~\cite{kashy19749} \\
     $\isotope[9][]{Be}(d,n)\isotope[10][]{B}$ &  5930.2 & 10  & Yong Sook \emph{et al.}~\cite{park1973spectroscopy} \\
     $\isotope[9][]{Be}(d,n)\isotope[10][]{B}$ &  5930.2 & 10  & Fife \emph{et al.}~\cite{fife1967excited} \\
    \hline
     \textbf{Average} & 5921.3 & 3.2 &  Without normalized \textit{Q} value\\
    \textbf{Average} & 5924.6 & 3.2 &  With normalized \textit{Q} value \\
     $\boldmath{\isotope[6][]{Li}(\alpha,\gamma)\isotope[10][]{B}}$ & 5927.8 &  $\pm 0.5$ (stat.) $\pm 2.4$ (syst.)   &  DRAGON 
     \\  \hline \hline
    
    \label{tab:energies}
    \end{tabular}

\end{table*}

\section{Discussion \& Conclusions}
\label{sec:discussion}

As this work demonstrates, such measurements can provide a test of 
the limits of the DRAGON angular acceptance, using a known 
resonance of the 
$\isotope[6][]{Li}(\alpha,\gamma)\isotope[10][]{B}$ reaction.
It is worth noting that such measurements can provide reliable
results only if they are coupled with detailed simulations
of the separator, which provide the important information
on the recoil transmission and the $\gamma$ ray detection efficiency.
The data analysis is also affected by our knowledge of
the $\gamma$ branching ratios and angular distributions $W(\theta)$,
as discussed in Section~\ref{sec:radiative}.
Accurate measurements of the $\gamma$ branching ratios using
high efficiency detectors are
desirable, but even if they are unknown,
the $\gamma$ ray detection efficiency can still be calculated using
a combination of the experimental data and simulations~\cite{ruiz2014recoil}.
The final result, however, might suffer from higher systematic uncertainty.

The results we obtained in this work are in excellent agreement with the only known 
measurement by Forsyth \emph{et al.}, showing that DRAGON
can measure resonance strengths of reactions with large 
recoil angular cones. We can now proceed with confidence 
to study a wide range of alpha--capture reactions
(see Table~\ref{tab:reactions}), previously
thought inaccessible with DRAGON due to recoil acceptance constraints.
This includes a planned measurement of the 
$\isotope[7][]{Be}(\alpha,\gamma)\isotope[11][]{C}$ reaction
at energies relevant to $\nu p$--process nucleosynthesis.

\section*{Acknowledgements}
 
The authors gratefully acknowledge the beam delivery and
ISAC operations groups at TRIUMF. The core operations of TRIUMF
are supported via a contribution from the federal government through the 
National Research Council of Canada, and the Government of British Columbia 
provides building capital funds. DRAGON and authors from McMaster University 
are receive funds from the National Sciences 
and Engineering Research Council of Canada (NSERC). 
A.P. thanks Greg Christian (St. Mary's University) for valuable 
discussions concerning the \texttt{GEANT} simulations, and
the anonymous reviewers whose comments/suggestions helped improve 
and clarify this manuscript.
Authors from the UK are supported by the Science and Technology 
Facilities Council. Authors from
Colorado School of Mines acknowledge
support from U.S. Department of Energy Office of Science DE-FG02-93ER40789 
grant.
This work benefited from discussions at the ``Nuclear Astrophysics at Rings and Recoil Separators" Workshop 
supported by the National Science Foundation under Grant No.
PHY-1430152 (JINA Center for the Evolution of the Elements).

%% References with bibTeX database:
\bibliographystyle{model1-num-names}
\bibliography{sample.bib}

\end{document}